%% file: main.tex
\documentclass[runningheads]{llncs}
\usepackage{graphicx}
\usepackage{hyperref}
\hypersetup{
    colorlinks=true,
    linkcolor=blue,
    filecolor=magenta,      
    urlcolor=blue,
    citecolor=blue
}
\usepackage[disable]{todonotes}
\usepackage{adjustbox}
\usepackage{algorithm}
\usepackage{algorithmicx}
\usepackage[noend]{algpseudocode}

\usepackage{multirow}
\usepackage{xspace}

\newcommand{\toolname}{{{\scshape{Ostinato}}}\xspace}
\newcommand{\holmes}{{{\scshape{Holmes}}}\xspace}

\usepackage{cite}
\usepackage{bbding}
\usepackage{pifont}
\usepackage{amssymb}
\newcommand{\xmark}{\ding{55}}

\usepackage{caption}
\usepackage{subcaption}
\usepackage{pbox}
\usepackage{makecell}

\usepackage{adjustbox}
\usepackage{amsmath}
\usepackage{dirtytalk}
\usepackage{mdwmath}
\usepackage{mdwtab}
\usepackage{titlesec}
\usepackage{enumitem}
\usepackage{mdframed}
\usepackage{lipsum}
\usepackage{tabularx}

\usepackage{colortbl}
\usepackage{hhline}
\usepackage{array}
\usepackage{color}

\makeatletter
\newcommand{\leqnomode}{\tagsleft@true\let\veqno\@@leqno}%
\newcommand{\reqnomode}{\tagsleft@false\let\veqno\@@eqno}%
\newcommand*{\compress}{\@minipagetrue}
\makeatother

\usepackage{titlesec}
\titlespacing*{\section}{0pt}{0.6\baselineskip}{0.6\baselineskip}
\titlespacing*{\subsection}{0pt}{0.6\baselineskip}{0.6\baselineskip}
\titlespacing*{\subsubsection}{0pt}{0.6\baselineskip}{0.6\baselineskip}
\begin{document}

\title{\toolname:  Cross-host Attack Correlation through  Attack Activity Similarity Detection}
\titlerunning{\toolname}
\author{Sutanu Kumar Ghosh, Kiavash Satvat, Rigel Gjomemo, and V.N.Venkatakrishnan\\ \email{\{sghosh34,ksatva2,rgjome1,venkat\}@uic.edu}}

\authorrunning{S.K.Ghosh, K.Satvat et al.}
%
\institute{University of Illinois Chicago, Chicago, IL 60607, USA}

%
\maketitle              
\vspace{-8mm}
\begin{abstract}
Modern attacks against enterprises often have multiple targets inside the enterprise network. Due to the large size of these networks and increasingly stealthy attacks, attacker activities spanning multiple hosts are extremely difficult to correlate during a threat-hunting effort. In this paper, we present a method for an efficient cross-host attack correlation across multiple hosts. Unlike previous works, our approach does not require lateral movement detection techniques or host-level modifications.  Instead, our approach relies on an observation that attackers have a few strategic mission objectives on every host that they infiltrate, and there exist only a handful of techniques for achieving those objectives. The central idea behind our approach involves comparing (OS agnostic) activities on different hosts and correlating the hosts that display the use of similar tactics, techniques, and procedures. We implement our approach in a tool called \toolname and successfully evaluate it in threat hunting scenarios involving DARPA-led red team engagements spanning 500 hosts and in another multi-host attack scenario. \toolname successfully detected 21 additional compromised hosts, which the underlying host-based detection system overlooked in activities spanning multiple days of the attack campaign. 
Additionally, \toolname successfully reduced alarms generated from the underlying detection system by more than 90\%, thus helping to mitigate the threat alert fatigue problem.


\end{abstract}

\input{introduction}
\input{problem_description}
\input{approach}
\input{evaluation}

\input{related_works}
\input{conclusion}

\bibliographystyle{splncs04}
\bibliography{bibliography}

\end{document}

%% file: introduction.tex
\section{Introduction}\label{sec:intro}

Modern advanced persistent threats (APT) often spread stealthily  across multiple hosts in their target enterprises. Detecting APT activities across multiple hosts inside such networks is very challenging.  Approaches that deal with this challenge are often {\em network-based}~\cite{cuppens2002alert,kruegel2004intrusion,sadoddin2006alert}.  They focus on finding  a strong presence of attack artifacts in network data (e.g., DDOS, botnets). However, modern APTs are increasingly stealthy and usually have a minimal footprint on network logs, and  are often characterized as ``slow and low''. Often, most of their actions occur {\em inside} hosts, while activities like scanning internal hosts or gaining access to new hosts happen over a long period of time. 

To be able to detect suspicious in-host activities, {\em host-based} solutions are needed. Current {\em host-based} approaches and Intrusion Detection Systems  (IDSes) \cite{hassan2019nodoze,milajerdi2019holmes,hossain2020combating} rely on audit logs  to detect attack activities represented as Indicators of Compromise (IOCs) or Tactics Techniques and Procedures (TTPs). However, they are focused on single-host detection and the alerts they raise are mostly about activities inside single hosts. To be able to deal with multi-host attacks, alerts raised on single hosts must be {\em correlated} with one another. 

One way to correlate alerts from multiple hosts involves understanding and detecting {\em lateral movement} tactics, techniques, and procedures (TTPs) employed by attackers\cite{rapid7-lm,splunk-lm,bowman2020detecting}. In particular, if a {\em lateral movement} TTP is detected, the two hosts involved in that TTP can be assumed to be victims of the same campaign. However, because such TTPs may be based on zero-day exploits or because of {\em threat alert fatigue} in human operators of Security Operation Centers \cite{awakelmfatigue,crowdlmfatigue}, they may not be detected and the alerts from multiple hosts may not be connected with one another.

 In this paper, we present \toolname,  a tool for efficient cross-host attack correlation across multiple hosts.  
 \toolname's design relies on the key observation that a specific APT group uses a finite (possibly large) set of tools during a campaign. In fact, according to MITRE's ATT\&CK page listing the cyber threat groups observed in the wild, a vast majority of those groups employ only a handful of techniques and procedures~\cite{mitreGroups,dfir}. 

 Based on this observation, we design an approach that compares (OS-agnostic) activities on different hosts and correlates those hosts that display similar suspicious techniques used to achieve similar objectives across those hosts. In particular, if similar tactics appear on two different hosts, then it is likely that the two hosts are victims of the same attack and they are, therefore, correlated. 

The main challenge in realizing this approach lies in defining an {\em activity similarity computation} method that can be applied independently of attack peculiarities and thus be used in a general setting in networks with a large number of hosts. To address this challenge, \toolname first models the attacker's techniques and the underlying operational procedures as {\em tagged provenance graphs}, which represent audit logs as graphs that are tagged with attacker-related procedures. Next, \toolname defines a novel  approximate graph similarity computation method that can be applied to the set of tagged provenance graphs in a pairwise fashion. The main contributions of the paper are as follows.

\noindent {\bf Graph- and Similarity-based Correlation.} We propose a novel approximate graph similarity-based alert correlation technique by addressing the (often overlooked) problem of determining when entities (e.g., processes, files, sockets, and its respective information flow) associated with alerts from different hosts are similar during an attack campaign. This is particularly useful for cross-host attack correlation involving hundreds of hosts in an enterprise network.

 \noindent
 {\bf Threat hunting application}. 
 The second contribution of this paper is the detection of compromised hosts through the correlation between detected attacker activities and other activities across multiple hosts using graph similarity. In this kind of application, \toolname can enhance the threat hunting capabilities of existing  Security Information and Event Management (SIEM) systems.

 \noindent
 {\bf Threat alert fatigue mitigation application}.  Threat alert fatigue is a common problem in Security Operation Centers (SOC), where human operators pour over hundreds of thousands of alerts generated by the network- and host-based systems. \toolname can  boost alert scores related to attack activities that are similar across multiple hosts and thus help reduce alert fatigue.

 Along with these contributions, we perform the experimental evaluation (Section \ref{sec:evaluation}) where we use two  different datasets collected from several red team engagements organized by DARPA. In the first dataset, the red teams performed various attacker activities across a network of 500 Windows hosts, resembling modern APTs. In this evaluation, the single host-based detection system either missed attacker activities having small footprints that evade its detection threshold or produced many false positives at lower thresholds. In turn, because of similarities among these activities \toolname was able to detect 21 additional hosts compromised by the attacker. We also created and evaluated an extensive dataset of more than 1000 graphs (of different sizes) generated by varying the detection threshold of the underlying IDS in each of the hosts from the same data.  In the second dataset, we further evaluated \toolname on a different attack scenario involving multiple hosts of different OSes and successfully correlated cross-host attacker activities.

This paper is structured as follows: Section 2 provides a high level description of the problem. Section 3 describes the approach and architecture, Section 4 contains the evaluation, Section 5 describes the related work and Section 6 the conclusion.

%% file: problem_description.tex
\section{Problem Description}\label{sec:overview}

In a multi-host system, one of the primary methods for expanding threat-hunting activities to new hosts relies on detection of {\em lateral movement} activities. In particular, if lateral movement events are seen on one host -- e.g., suspicious traffic to a remote Windows SMB host --  SOC operators may decide to escalate the alerts on both those hosts to more scrutiny. However, this strategy relies on {\em known} lateral movement indicators, and it may not always work if those indicators are missing, incorrectly modeled, if attackers modify their tactics so that they do not match known indicators, or if they move laterally via existing benign network communications (living off the land).    

{\em Alert fatigue} is another cause for failing to process lateral movement indicators. In modern SOC centers, with thousands of hosts and hundreds of thousands of alerts (the majority of which are false positives), without a (relatively) strong signal about lateral movement or initial compromise on a host, it is counterproductive to escalate alerts to a higher level of scrutiny. Operations in such centers are finely tuned to deal with {\em alert fatigue}, and almost every system incorporates techniques, filtering mechanisms, and knobs that adjust the signals to forward to human operators~\cite{crowdscore}. Setting such filters to low values ensures reducing false negatives at the cost of having more false positives. Setting them at higher values reduces  false positives but can potentially miss true positives. As a result, legitimate lateral movement indicators may be ignored by analysts.

 \begin{figure}[t]
    \centering
    \includegraphics[width=0.8\textwidth]{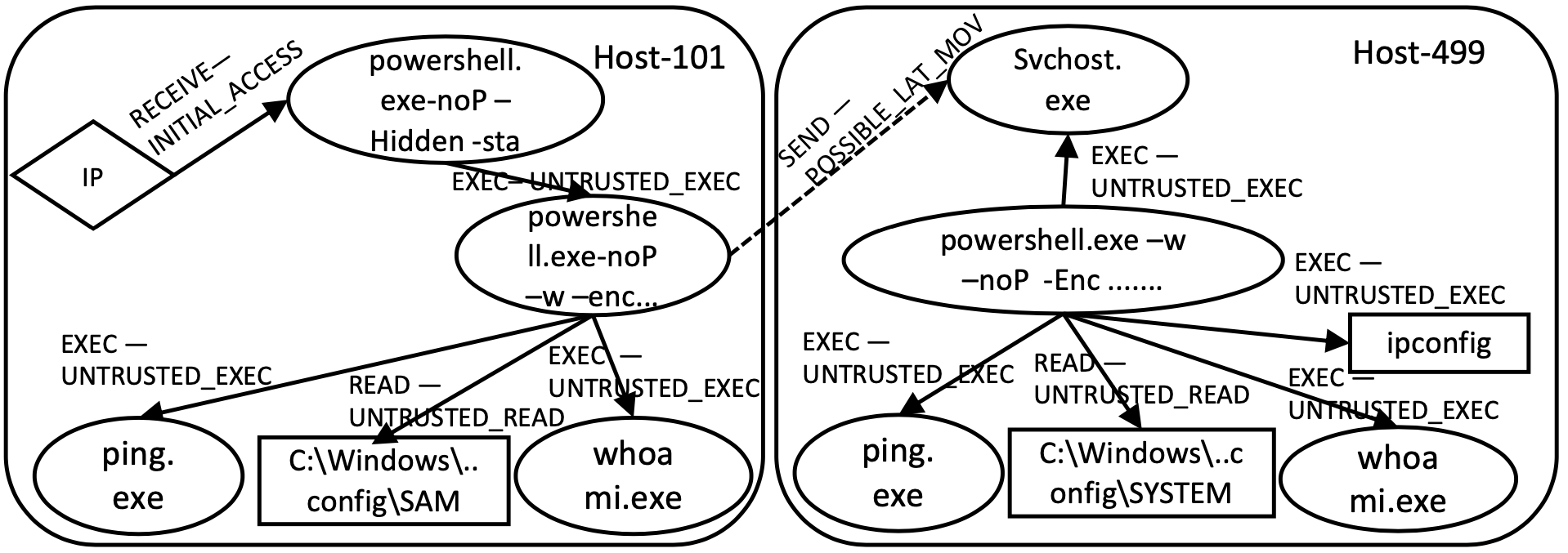}
    \caption{Example of alarms raised by underlying IDS in multiple hosts in the form of tagged provenance graph}
    \label{fig:running-eg-alarms}
\end{figure}

 To illustrate the problem, consider the following running example (Figure \ref{fig:running-eg-alarms}). An attacker obtains an initial foothold in a host (Host-101) inside a network of several hundred hosts. There (s)he performs several actions on the compromised host using powershell commands. These include pinging other hosts, monitoring running processes, reading sensitive system files, password hashes, and so on. The host-based intrusion detection system (IDS) raises alerts for some of these activities. These alerts are related to events in the audit logs that have some suspicious connotations. However, because they are similar to benign activities, they do not pass the threshold needed to be forwarded to a human operator or, if they do, they may appear together with many other false positive alerts, and thus be missed by human operators. Next, the attacker uses  compromised passwords from Host-101 to gain access to another host (Host-499), which is usually accessed remotely by a benign user from Host-101. Because the attacker is mimicking a benign activity, this lateral movement remains undetected. The attacker  performs similar actions in the new host including pinging other hosts, monitoring daily activities, sensitive file systems, and password hashes. The IDS running on Host-499 is identical to that on Host-101 and raises similar alerts. However, because the connection between Host-101 and Host 499 is considered benign, alerts are again missed. As a result, the attack is not detected or ignored. 
 
\noindent
\textbf{Problem Statement}.
Our problem statement is as follows: \textit{How can we correlate alerts across hosts without relying on lateral movement detection in a network with hundreds of hosts? How can we obtain an additional suspicious signal related to correlation for SOC operators? }
Our key observation to solve this problem is that those attacker activities that are observable in audit logs are a manifestation of the attackers' overall goals (i.e., kill-chain steps) and related techniques \cite{mandiant}. Often, these goals overlap across hosts. For instance, for every host that is compromised, there {\bf must} exist an {\em initial access} step. Often, initial access is followed by a {\em discovery} step, where the attacker explores the newly compromised host. To spread to a new host, the attacker {\bf must} perform {\em lateral movement}. To be able to maintain their presence in the hosts for a long time, the attackers {\bf must} execute some form of {\em persistence}. 

Another {\em key insight} at the basis of our solution is that the tactics and procedures available to carry out these common goals in multiple hosts during an APT campaign are not infinite but are limited in number. The attackers must, therefore, execute similar procedures in several hosts. In particular, by the pigeonhole principle, the larger the number of compromised hosts in a network, the more likely it is that similar activities are carried on those hosts. While it is certainly possible that attackers use different procedures for the same goals on different hosts, this would significantly raise the bar of difficulty for the attackers as this would require exploiting several vulnerabilities and increasing the chances of being detected by the underlying IDSes. Several research and survey papers, in fact, confirm the validity of our insight in the wild~\cite{apt_github,lm5tools2,mitrelateraltooltransfer}. As evidenced by some observations \cite{hajizadeh2018probability}, creating novel TTPs often requires significant resources and motivation from attackers.

The main challenge, in correlating cross-host alerts resides in producing a  similarity definition for alerts that is general enough to be used across multiple hosts in a network. In particular, because each host can execute processes in many different ways, we must be able to capture the similarity between processes' behaviors inside different hosts. We solve this challenge with \toolname, a system that can detect similar behaviors present in the alerts generated by IDS-es, and create additional alerts. \toolname's goal is to be used as companion to existing IDS-es and provide an additional signal for attack detection.

%% file: approach.tex
\section{Approach and Architecture}\label{sec:approach}
\noindent
\textbf{Threat Model.}
We assume that the attacker is able to initially compromise a host and, starting from that host, spread to other hosts inside the network, either via relying on vulnerable processes or by using existing tools from the compromised host (e.g., remote desktop services, SSH, etc). We assume that there is an intrusion detection system in each host that is generating alerts, which may detect part of the attacker's activities. However, these alerts are also  buried inside a large number of false positive alerts. Similar to prior research in this area, we also assume that the audit logs data are trustworthy and not modified by the attackers. We also assume that the alerts are derived from existing audit logs systems (ETW, Auditd) and that they contain the system calls generated by the running processes and the process invocations with their command line arguments. We represent the information in these logs and alerts as tagged provenance graphs.

\subsection{Tagged Provenance Graphs}

Provenance graphs ~\cite{lee2013loggc, krishnan2010trail, hossain2020combating,satvat2021extractor, king2003backtracking, milajerdi2019holmes, milajerdi2019poirot,hassan2019nodoze} are well-known, widely-popular representation of audit logs, where nodes represent system entities  (processes, files, registry entries, sockets) labeled by the entity names or paths together with command line arguments (in the case of processes) and directed edges represent system events and system calls (and are labeled by the system call name, e.g., read, write, fork, mmap) that connect those entities. 

To represent both the high-level attacker goals and their low-level operational details at the same plane, \toolname enhances {\em provenance graphs} with tags (additional labels) on the edges of the graph representing semantic level details (attacker goals, tactics names, and others).  This enhancement is done by \toolname based on the graph and its respective alarms generated by the underlying IDS where the edge names are augmented with additional information before storing them in a common database. Examples of such graphs are shown in Figure  \ref{fig:running-eg-alarms}. The nodes represent different system entities, while  edges are labeled by both {\em system call labels} ({\tt exec}, {\tt remove}) and  {\em suspiciousness labels}, which capture the attacker's goal ({\tt Untrusted Exec}, {\tt Untrusted\_Remove}), as a TTP \cite{attAndck} name. This novel enhanced representation, which we call {\em tagged provenance graphs}, allows us to represent alerts including both  system behavior and attacker goals and include these high level details in the search for similar alerts.

 Using {\em tagged provenance graphs}, \toolname models the process of alert correlation across different hosts as a search for similar tagged provenance graphs representing those alerts. In particular, \toolname first determines \textit{node similarity} between nodes in different graphs. Next, it uses the edge labels and tags to determine edge similarity, and  finally combines the nodes and edge similarity values into an overall similarity score for a pair of tagged provenance graphs. 
 
\begin{figure*}[!t]
    \centering
    \includegraphics[width=13cm,height=6cm]{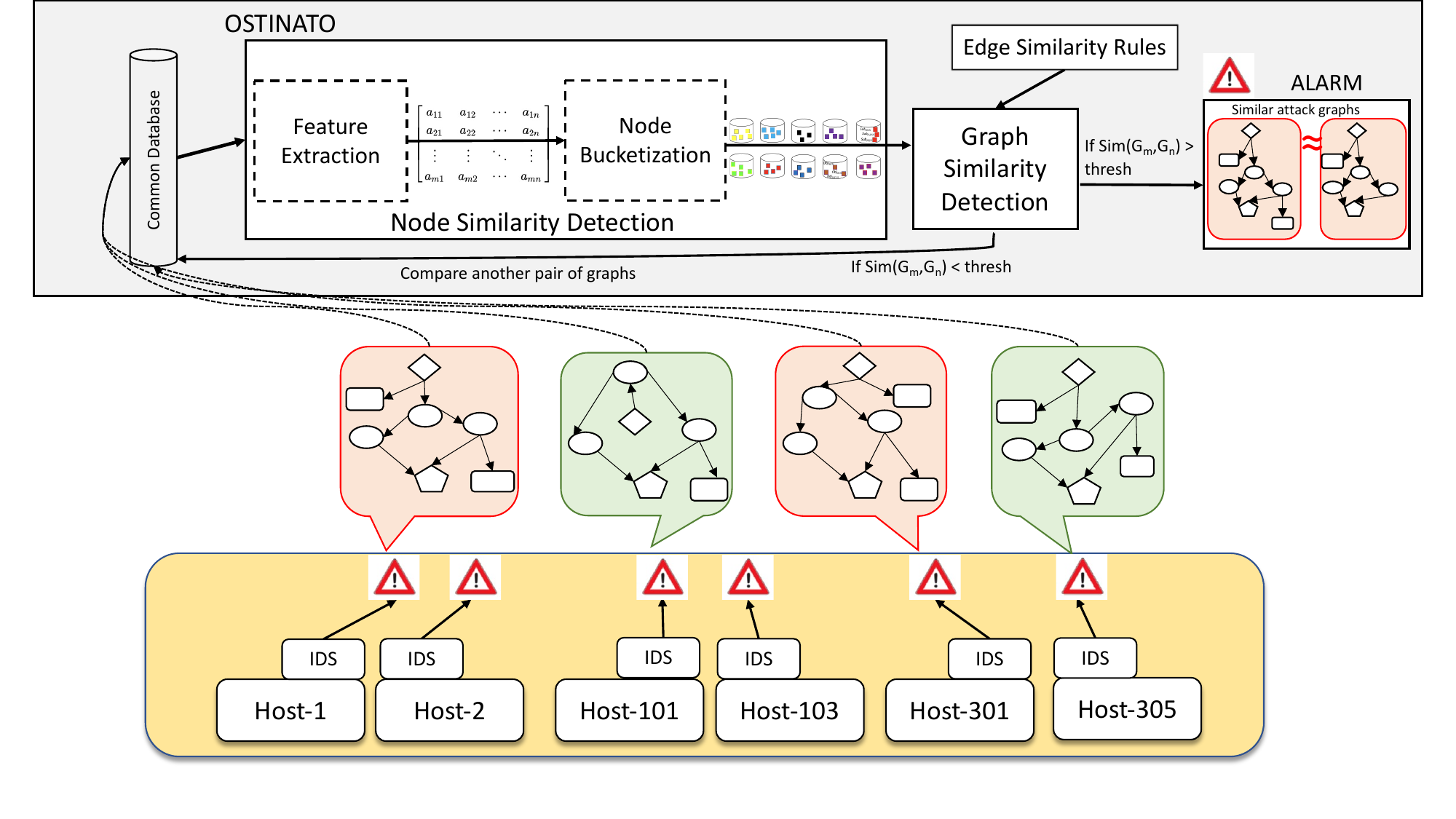}
    \vspace{-8mm}
    \caption{\toolname Architecture.  }
    \label{architecture fig}
    \vspace{-2mm}
\end{figure*}

 \toolname's architecture is shown at the top half of Figure \ref{architecture fig}.
 At the bottom of the figure, we show the hosts of an enterprise network. The IDSes inside each of these hosts produce alerts that are next transformed into tagged provenance graphs and stored in a central database by \toolname. It serves as a companion to these IDSes and utilizes the respective tagged provenance graphs. Each of these tagged provenance graphs in the central database are processed by the first phase of our approach, {\em Node Similarity Detection} (Sec. \ref{subsec:nodesimi}), which is responsible for grouping different nodes from different graphs into {\em buckets} containing similar nodes. Next, the {\em Graph Similarity Detection} step uses these buckets and a set of edge label similarity rules to compute the final similarity value among the tagged provenance graphs (Sec. \ref{subsec:edge_sim}, \ref{subsec:algo}).  Finally, if the similarity value crosses a specific threshold, \toolname raises an alert. 
The  details of each of these stages along with their challenges  are explained in the remainder of this section.

\subsection{Identifying Similar nodes}
\label{subsec:nodesimi}
There is a large body of work dedicated to computing graph similarity and related problems, including graph isomorphism~\cite{mckay2014practical}, iterative (or structural) methods~\cite{simRank}, graph pattern matching~\cite{gallagher2006matching}. At the foundation of these algorithms, often there is an {\em assumption that an initial mapping between nodes on different graphs already exists}. Such mapping informs these algorithms on which nodes in one graph are similar or match which nodes in another graph. They often assume that nodes have simple labels (e.g., single letters of the alphabet), which can be trivially used to provide an initial similarity measure between nodes. 
In our setting, however, nodes in the provenance graphs represent different entities, including processes, files, sockets, etc. An initial mapping that can inform about similar nodes across hosts does not exist. Therefore, one question at the core of our problem is: how can we produce such mapping? When can we claim that, for instance, two processes from two different hosts are the same or similar? 
Is, for instance, a {\em PowerShell} process from one host similar to a {\em PowerShell} process from another host?
To answer this question, we focus on two aspects of nodes in the graphs: {\em node label content} and {\em approximate node behavior}.

\noindent
\textbf{Node Label Content.} Node labels consist of text extracted from audit logs information. They typically contain identifiers, e.g., the names and paths of the entities,  command line invocation (for processes), flags, and other entity definitions. Because these entities are in different hosts, such labels may not be the same, even for processes often presumed to be similar. For instance, exact string comparison would be unable to identify the two nodes from two different graphs \path{C:\Windows\System32\WindowsPowerShell\v1.0\PowerShell} \texttt{ -noP -w 1} and \path{C:\Windows\System32\WindowsPowerShell\v1.0\PowerShell.exe} \texttt{ -NoP -NonI -w } as similar when actually they behave similarly in the host level. Furthermore, attackers may also invoke processes differently by using  different order of similar command-line arguments and other means. An effective node similarity computation method must take into consideration all these factors.

\noindent
\textbf{Approximate Node Behavior.} Different variations of string comparison would also not result in accurate node similarity. For instance, one can extract only the file name from the path of an image before performing the string comparison so that different directory structures do not interfere with the matching. In this case, labels like \path{C:\Windows\System32\WindowsPowerShell\v1.0\PowerShell.exe} and \path{PowerShell.exe} would match. While such a solution might perform better than string comparison in certain cases, it would perform poorly for complex processes such as PowerShell, python that act as interpreters. These processes can exhibit multiple different behaviors depending on their input and command-line arguments and cannot be assumed to be similar only because they have the same name.

\noindent
\textbf{Solution}. In our solution, we consider the node labels as textual representations of the nodes' behavior and use a notion of text similarity to determine node similarity. Beyond the two naive string-based approaches that we mentioned, there are several other approaches used for dealing with text similarity like Bag of Words \cite{joachims1998text}, Word2Vec \cite{mikolov2013efficient}, or ConceptNet \cite{liu2004conceptnet}. These approaches, however, incorporate concepts derived from natural text, such as synonyms, and cannot be directly used in our problem where the text is composed of processes, file names, and command-line options.  In practice, a good solution should: 1) be applicable to the domain of terms appearing in audit logs and not rely or depend on assumptions of natural text, 2) use general intuitions about anomalies in data related to attacker activities. 

We first consider each node label as a text document and use the {\em Term Frequency-Inverse Document Frequency (TF-IDF)} method \cite{joachims1996probabilistic} to create a feature matrix that captures the presence of words inside the nodes and their importance. TF-IDF makes no assumptions on the kind of text it acts on, and it allows evaluating how relevant a phrase is to a document in a group of documents as a statistical measure. Next, we use Locality-Sensitive Hashing (LSH) on the feature matrix \cite{shrivastava2014defense} to create {\em similarity buckets}, where nodes in the same bucket are similar to each other.  We describe these two steps next.

\subsubsection{Feature Extraction:} \label{subsec:tfidf}
The first step in our approach  creates a matrix representation of the node labels in the tagged provenance graphs, which can be used in the next step of the approach. This matrix, which we call {\em feature matrix}  is built by considering the node labels as text documents and applying the TF-IDF measure over them. 
TF-IDF is a measure for determining how relevant a word is inside a set of documents~\cite{joachims1996probabilistic}. 
The TF part represents the number of times a term appears in a single document (node). The IDF part, on the other hand, represents the informativeness of a term. In particular, a term that appears more frequently in all the nodes is expected to be less informative compared to one that rarely appears in those nodes.
In our approach, we define as `terms' the words inside the node labels separated by white spaces, and we consider each node as a separate document. In particular, each node (i.e., subject or object) is a document $d_i$: $d_{i} = \{t_{1},t_{2},t_{3},...,t_{n}\}$. We create two sets of such documents $D_{sub}$, representing all subject nodes (processes), and $D_{obj}$ representing all object nodes (registry, files, IP). In the following details, we describe only the steps related to $D_{sub}$ for space reasons. Steps for $D_{obj}$ are identical. 
In the first step, we calculate the TF-IDF score for each term appearing in the documents in $D_{sub}$. This score is the product of the two features: The term frequency of term t in document $d_i$ is shown in Equation \ref{eqn1} where the numerator is the number of times  $t$ appears in  $d_i$, and the denominator represents the total number of terms ($t^{\prime}$) in $d_i$. The inverse document frequency of the term is shown in Equation \ref{eqn2} where the numerator is the total number of documents (subject nodes), and the denominator represents the number of documents (subject nodes) that contain term $t$.

    {\centering\compress
\begin{tabularx}{\linewidth}{>{\leqnomode}XX}
\vspace{-6mm}
\begin{equation}
\label{eqn1}
\begin{aligned}
      \operatorname{tf}(t, d_i)=\frac{f_{t,d_i} }{\sum_{t^{\prime} \in d_i} f_{t^{\prime}, d_i}}
\end{aligned}
\end{equation}
 &
 \vspace{-6mm}
 \begin{equation}
 \label{eqn2}
\begin{aligned}
\operatorname{idf}(t, D_{sub})=\log \frac{|D_{sub}|}{|\{d_i \in D_{sub}: t \in d_i\}|}
\end{aligned}
\end{equation}

\end{tabularx} \vspace{-\baselineskip}}

After calculating the TF-IDF score for all terms in the documents in $D_{sub}$, for each $d_{i} \in D_{sub}$, we calculate the median score $\hat{\mu}$  of the TF-IDF values of its terms. 
Next, we build a matrix $M_{sub}$ with dimensions $m \times n$ where $m$ equals the number of documents $|D_{sub}|$ and $n$ equals the total number of terms appearing in the documents in $D_{sub}$. A row in the matrix represents a document, while a column represents a term. 
If term $t$ exists in $d_{i}$:
\vspace{-7mm}
\begin{figure*}[h]
    \centering
    \includegraphics[width=0.4\textwidth]{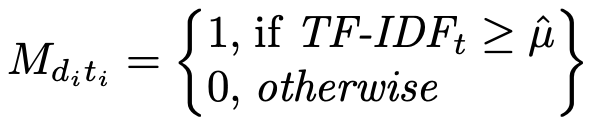}
    \vspace{-7mm}
\end{figure*}

\begin{figure}[t]
    \centering
    \includegraphics[width=1\textwidth]{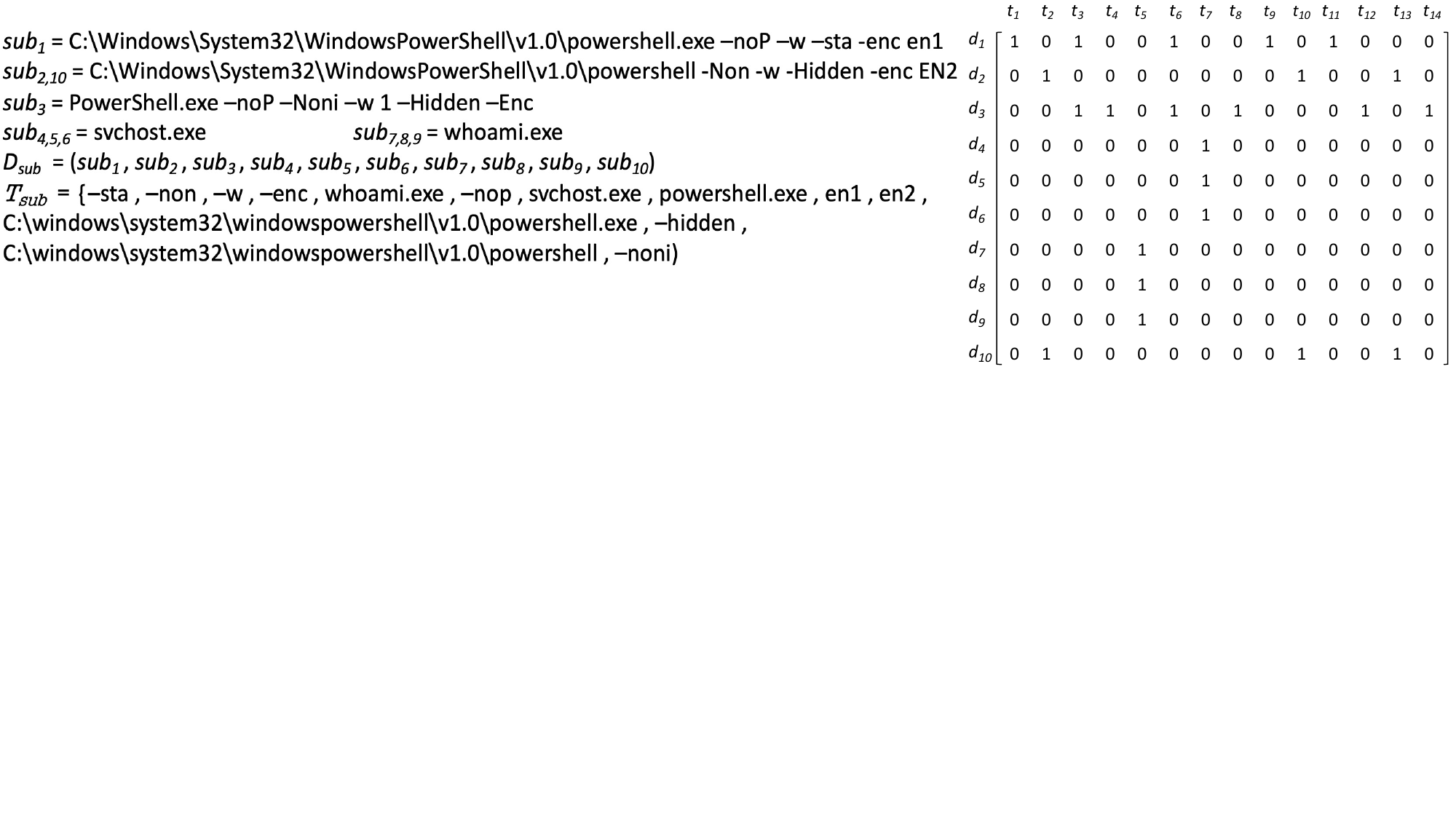}
    \vspace{-42mm}
    \caption{Example feature matrix. The rows represent node labels. The columns represent words and the cells represent words with TF-IDF higher than median TF-IDF of the corresponding document.}
    \label{fig:tfidf-running}
\end{figure}

\noindent The $M_{sub}$ matrix represents the relevance of each term inside each document. The intuition behind this matrix is that we want to keep track of the terms which have appeared in the nodes and, at the same time, are important ($\geq \hat{\mu}$). This means two $1$ in two different nodes (i.e., rows) in the same column indicate the presence of an important (relatively rare) term which is $\geq \hat{\mu}$ in both nodes. On the other hand, the terms that are non-relevant and therefore less informative and important ($< \hat{\mu}$) will be represented as zeros.  Figure \ref{fig:tfidf-running} depicts an example of a $M_{sub}$ matrix. The 10 subject nodes in the tagged provenance graphs are shown at the top of the figure. These correspond to 10 rows in the matrix. Each column in the matrix represents one of the words presented in the node labels. The matrix cells represent the presence (1) or absence (0) of a TF-IDF value that is larger than the median TF-IDF in each row.

\subsubsection{Node Bucketization.} \label{subsec:lsh}

The next step in our approach is to determine the (approximate) similarity between nodes, represented by the rows of the feature matrix $M_{sub}$ (and $M_{obj}$ for the object nodes). In particular, we want to cluster similar nodes into similar `buckets'. To do this, we   compute the similarity between each pair of rows in the TF-IDF matrix $M$ by using the Jaccard similarity measure among them~\cite{niwattanakul2013using}. 
This measure is calculated using both the number of elements that the two rows share and the number of elements they do not.  However, using this measure directly on the rows of matrix $M$ would present some scalability issues. For a matrix of m rows and n columns, the time complexity of these comparisons is $O(nm^2)$. Given that there could be millions of nodes and several (hundreds in some cases)  terms, this method would be computationally expensive. 

To deal with this issue, we use a version of Locality-Sensitive Hashing (LSH) with Minhash~\cite{broder2000min}.
LSH traditionally employs shingling, which breaks down  large documents into sequences of length $k$ of characters called \textit{k}-shingles. Used traditionally for detecting near-duplicate documents (e.g., plagiarism detection), LSH methods hash data records into buckets such that records similar to each other are placed in the same bucket with a high probability. In contrast, records distant from each other are likely to be placed in separate buckets.

In \toolname, we adapt LSH to solve our problem by using the TF-IDF feature matrix $M$ rows instead of k-shingles as input. Because the TF-IDF feature matrix encodes a semantic representation of  documents (i.e., nodes) that \textit{k}-shingles do not have, we believe this is a better approach than using only the LSH method over the documents. In particular, its {\em Minhash} function, can project high-dimensional binary vectors like $M_{sub}$ to a low-dimensional vector of integers $H$ by reducing the sparseness of the former. This transformation has the property that if the Jaccard index, $J(d_i, d_j)$ between two rows of $M_{sub}$ is high, then the probability value $Pr(H(d_i)==H(d_j))$ is also high. After creating the signature matrix $H_M$, we calculate pairwise row similarities  using the  formula:

\vspace{-1mm}
$$Sim(H(d_i), H(d_j)) = \frac{|H(d_i) \cap H(d_j)|}{D}$$ 
where the numerator is the size of the row intersection operator (over integers) and the denominator is the size of the rows. The value of this similarity is between 0 and 1. We finally place two nodes in the same bucket if their corresponding similarity is above a threshold $J_T$. This threshold is specific to the kind of data a system produces and can be tweaked by a domain expert based on their knowledge of the hosts and audit logs they produce.

An evasion technique that  attackers may try to use is to change the number of command line arguments in order to have two subject nodes in different buckets. This technique, however, is not likely to be successful for several reasons. To carry this out, the attacker has to include the command line arguments that carry out the objectives in the two subject nodes. To be able to place the nodes in different buckets, the attacker must change the values of the corresponding terms so that the terms' TF-IDF values are below the median in one node (so as to be represented as a 0 in the feature matrix) and above the median in the other node (so as to be represented as a 1 in the feature matrix). We point out that, due to the presence of the IDF, these median values cannot be controlled by the attacker but are a parameter of the system as a whole. Thus, if for instance, an attacker tries to modify the TF term by adding more values in the command line, they would  also inherently change the IDF term. This effectively raises the bar of difficulty for the attacker.

\begin{table}[t]
\centering
\begin{adjustbox}{width=0.9\linewidth}
\begin{tabular}{|l|l|} 
\hline
\textbf{Information Flow Similarity }       & \textbf{Prerequisites}                                                                                                                                                                                                                                                                                                                                         \\ 
\hline
$E_i \equiv E_j$                 &system call labels $E_i$ and $E_j$ are the same                                                                                                                                                                                                   \\ 
\hline
Load $\equiv$ Exec               & for all cases                                                                                                                                                                                                                   \\ 
\hline
Fork $\equiv$ Exec               & for all cases                                                                                                                                                                                                                   \\ 
\hline
Write $\equiv$ Create            & for all cases                                                                                                                                                                                                                   \\ 
\hline
Read $\equiv$ Exec               & \begin{tabular}[c]{@{}l@{}}$(Read.sub \approx $ PowerShell$) \wedge (Exec.sub \approx$ PowerShell$)$ \\ $\wedge$ $Read.obj$ $\approx$ \{.ps1, .psd1, .psm1\} \\$\wedge$ $Exec.obj \approx$ \{.ps1, .psd1, .psm1\}\end{tabular}  \\ 
\hline
TaskStart $\equiv$ ProcessCreate & \begin{tabular}[c]{@{}l@{}}$ (S(TaskStart) \in$ \{Untrusted\_Exec\} \\$\wedge S(OpenProcess) \in$ \{Untrusted\_Exec\}) \end{tabular}                                                                                            \\ 
\hline
Read $\equiv$ Load               & if $(Read.obj \in$ \{shared\_objects\} $\wedge$~$Load.obj \in$ \{shared\_objects\}$)$                                                                                                                                               \\
\hline
\end{tabular}
\end{adjustbox}
\vspace{+3mm}
\caption{Edge Label Similarity Rules. {\em S(name)} denotes the suspiciousness label, {\em Label.sub} is the subject and {\em Label.obj} is the object. $\approx$ denotes string containment. }
\label{tab:edge similarity rules}
\end{table}

\subsection{Edge Label Similarity} \label{subsec:edge_sim}

The edge labels can be very valuable in determining the similarity among tagged provenance graphs. In particular, {\em system call} labels can inform us about activity similarity at OS level, while {\em suspiciousness} labels carry much more meaningful information about attackers' goals. To capture edge label similarity, we incorporate several matching rules in \toolname. Given the finite number of suspiciousness and system call labels, this task does not need to be automated and can take advantage of domain knowledge. The edge label similarity rules that are used in \toolname are shown in Table \ref{tab:edge similarity rules}. The first column shows the similarity between edges using system call label names, while the second column shows the prerequisites that must be met for two edges to be considered similar. We also require that the suspiciousness labels are the same for all edge pairs (we do not show this in the table for space reasons). For instance, two edges with {\em exec} system calls labels are considered similar only if their suspiciousness labels are also the same (e.g., {\em Untrusted\_Exec}). 
In Table \ref{tab:edge similarity rules} the first row represents the trivial cases where both types of labels are the same (e.g., {\em read} and {\em read)}. The following rows represent cases where edges with different labels can be considered similar. For instance, the fifths row represents a rule that states that a {\tt read} in host $i$ is equivalent to an {\it exec} in host $j$ if either subject contains ($\approx$) `PowerShell' and if the suspiciousness label of either edge is different from {\it Initial\_Compromise}. This rule captures the duality of PowerShell scripts, which can be both read and execute. 
We point out that this table only deals with similarity among edge labels without considering the nodes. In other words, the table only captures information flow similarity. To fully evaluate if an event is similar to another, we also need to make sure that the nodes connected by that edge are similar to one another. We provide the details about this procedure in the next section.

\begin{algorithm}[!t]
    \caption{: Graph Similarity Algorithm. \label{alg:simi det}}

    \begin{algorithmic}[1]
    \Function{Similarity}{}
    \State \textbf{Input:} $G_l, G_s$, Buckets map $B: nodes \rightarrow  buckets$, Edge label similarity rules $E_L$,  MPS = $\{(N_{s},N_{l}) | N_{s} \in G_s \wedge N_{l} \in G_l \wedge B(N_{s}) = B(N_{l}) \}$, $Len_{MPS} = |MPS|$
    \State \textbf{Output:} Final\_Sim($G_l, G_s$) 
    
    \For{$(N_{s},N_{l}) \in MPS $}
        \State MPS = MPS \textbackslash \ $(N_s, N_l)$
        \State Total\_Acc += Parallel\_BFS$(N_{s},N_{l})$
    \EndFor
    \State Final\_Sim$(G_l, G_s) =   \frac{Total\_Acc}{Len_{MPS}}$ 

    \EndFunction
    
    \Function{Parallel\_BFS$(N_{s},N_{l})$}{}
    \State Sim = 0
    \State Enqueue$(N_s ,Q_s ) $; Enqueue$(N_l ,Q_l ) $
    \While {$(Q_s \neq \varnothing \wedge Q_l \neq \varnothing )$}
        \State $N_s$ = Dequeue($Q_s$); $N_l$ = Dequeue($Q_l$)
        \State MPS = MPS \textbackslash \ $(N_s, N_l)$
        \State $NN_s = \{V | (N_s ,V) \in E(Gs) \vee (V, N_s ) \in E(G_s )\}$
        \State $NN_l = \{V | (N_l ,V) \in E(G_l ) \vee (V, N_l ) \in E(G_l )\}$
        \For{$v_1 \in NN_s $}
            \For{$v_2 \in NN_l $}
                \If{$(v_1 , v_2) \in MPS $}
                    \State Enqueue$(v_1 , Q_s)$; Enqueue$(v_2 , Q_l)$
                    \If{ $E_L (N_s , v_1) == E_L (N_l , v_2) $ }
                        \State Sim+= $W_1$
                    \Else
                        \State Sim+= $W_2$
                    \EndIf
                \Else
                    \If{ $E_L (N_s , v_1) == E_L (N_l , v_2) $ }
                        \State Sim+= $W_3$
                    \EndIf
                \EndIf
            \EndFor
        \EndFor
        \Return Sim
    \EndWhile
    \EndFunction
    \end{algorithmic}
\end{algorithm}


\subsection{Graph Similarity Detection} \label{subsec:algo}
The final step of \toolname, is to determine whether two tagged provenance graphs belonging to two different hosts are similar or not. These graphs, however, can: 1) have widely different sizes, depending on the number of suspicious activities detected in each host, 2) be composed of different activities that may or may not be similar. To determine the final similarity score between two tagged provenance graphs, we use Algorithm \ref{alg:simi det}, which performs in parallel two modified breadth first searches over the two tagged provenance graphs while updating a similarity score value during the traversal. This algorithm uses both the bucket information representing the node mappings and the edge label similarity rules to determine whether an initial attack graph is similar to another graph (or a set of graphs) in comparison to the attack behavior and the structure of the graph. 

Algorithm \ref{alg:simi det} takes in input the two tagged provenance graphs, the edge label similarity rules $E_L$ (Table \ref{tab:edge similarity rules}), and  Matched Pairs Set (MPS), which is the set of pairs of nodes from the two graphs that are in the same bucket. The algorithm chooses one such pair of nodes and performs a breadth first search traversal on each graph  using those nodes as roots. Before the traversal, it removes that pair of nodes from the set, so that it does not traverse them a second time later. During the traversal, it only follows the nodes of the two neighborhoods that are in the same bucket (lines 11-16). At any iteration of the loop in line 16,  considers three cases of similarity, to which it assigns three different weights: 1) $W_1$ corresponding to complete edge matching (nodes and edge labels), 2) $W_2$ corresponding to the two nodes matching but the edge labels being different (E.g., firefox writes to a file in one host and firefox reads from the same file in another host), 3) $W_3$ corresponding to the case where the subject node and edge labels match but the object names do not match.  This approach works across hosts with different OS because even though the names of  processes are varied across different OSes, the malicious behaviour and its usage would place the nodes into the respective similar bucket. 
We use different weights in order to take into account the differences in the number of buckets of subjects and of objects discussed earlier (see end of Section \ref{subsec:lsh}). The weights we used in our evaluation for $W_1, W_2$ and $W_3$ are 1, 0.2 and 0.8 respectively. From our evaluation, we conclude that these values can be generalized for different OSes or platforms. Additionally, the value of these weights can be customized further by analysts to look for specific nodes during forensic analysis or threat hunting.

After the final similarity score between two graphs is determined in line 7, we  raise an alert if it is higher than a predefined similarity threshold. The value of this threshold depends on several factors, including the systems and the filtering actions of the local IDS detectors. We include a discussion about this threshold and others in the Evaluation.

%% file: evaluation.tex
\section{Evaluation}
\label{sec:evaluation}

This section evaluates \toolname by two different experiments using different datasets generated by DARPA  red team exercises. The first experiment is part of large-scale  3-day long red team exercise \cite{dataset} in an environment containing 500 Windows hosts in which the major attacker activities were concentrated in the first two days. The details of this experiment are discussed in Section \ref{efficacy_sec}. We further evaluate \toolname on a second experiment which contains two separate multi-host attack campaigns involving hosts with different OSes\cite{tc5}. 

We deployed \toolname on a desktop with  Intel Xeon W CPU @ 3.2GHz and 32 GB memory running macOS Big Sur. As a local IDS, we used \holmes,  which we obtained from its developers~\cite{milajerdi2019holmes}. \holmes uses rules of {\em connected} TTPs to detect attacks unfolding inside a single host. Its final output consists of  provenance graphs representing the activities detected as TTPs. These graphs are next sent to \toolname, which determines similarities among them. 

\noindent
\textbf{Results Summary.}
We performed our experiments in a {\em threat hunting} scenario where, given some attacker activities in one host, we use \toolname to find similar activities in other hosts. Thanks to this kind of search, \toolname was able to uncover attacks in 14 more hosts than \holmes on the first day (detailed description in \ref{efficacy_sec}) and 7 more (21 in total) on the second day. This is due to the lighter footprint of the attacks on the additional hosts, which fall under \holmes' detection threshold. In fact, to make \holmes detect the same attacks as \toolname on those additional hosts, we had to lower \holmes' detection threshold significantly, producing  several hundreds of alerts and false positives. 

\subsection{\toolname Efficacy}

\noindent
\textbf{Dataset Overview}\label{setup&dataset}
We evaluate \toolname over two datasets: first, OpTC-NCR2 a large dataset \cite{dataset} of audit logs produced as part of DARPA's CHASE program. The dataset was collected over a period of two days on 500 hosts. During these days, a red team performed several APT-like attacks on 24 of those hosts. Benign activities were generated both manually and by running scripts. 
The second dataset was collected as part of DARPA's Transparent Computing (TC) program \cite{tc5}. During this engagement, the attackers replicated APT-type scenarios across multiple hosts on different platforms.

\noindent
\textbf{Ground Truth.} The  data are accompanied by PDF documents written by the red team describing the attackers' activities performed on each host. The ground truth was built from these descriptions and the process ids contained in those descriptions. In particular, if  a tagged provenance graph contains one or more of those process id-s it is considered as an {\em attack graph}.
In addition, we build a ground truth of pairs of similar attack graphs manually.

\noindent\textbf{Detailed Results.}
Table \hyperlink{day2}{2} shows \toolname's results for the first two days of the OpTC-NCR2 dataset. The left table (a) contains the results of the first day, while the right table the results of the second day. The tables contain pair-wise similarity scores among tagged provenance graphs that were a part of the attackers' activities and the maximum similarity score (Column $B_{max}$), and mean similarity score (Column $BM_1$) between each graph that represents attacker activities and the other provenance graphs that represent benign activities. In Table \hyperlink{day2}{2}(b), $G_a$, $G_b$, and $G_c$ represent 3 tagged provenance graphs generated by \holmes (and enhanced by \toolname) in its default optimal detection threshold setting, which produces true positives and a low number of false positives. These were present on only one host, hence comparisons among them are not calculated. The rest of the tagged provenance graphs from 7 distinct hosts $(G_{d}-G_{j})$ represent activities with a smaller footprint, which were not detected as attacks by \holmes in its default detection threshold. In our experiments, we reduced \holmes's detection threshold obtaining a total of 689 more graphs from 500 hosts. Using \toolname, we identified 7 (from 7 distinct hosts) out of 689 graphs that were similar to the initial 3 tagged provenance  graphs as part of the attacker's activities, while the rest were false positives. In these hosts, the attack's footprint was smaller because the attackers performed only a small number of malicious activities like running  some PowerShell scripts in some hosts or communicating to an untrusted C2 server in other hosts. There were several benign graphs generated in those 7 affected hosts, for which \toolname generated low similarity scores as per expectations. \toolname was able to successfully correlate attacker activities found in the initial attack graphs ($G_a$, $G_b$, $G_c$) among hundreds of other graphs. 

\label{efficacy_sec}
\begin{figure*}[t]
    \centering
    \vspace{-2mm}
    \includegraphics[width=1\textwidth]{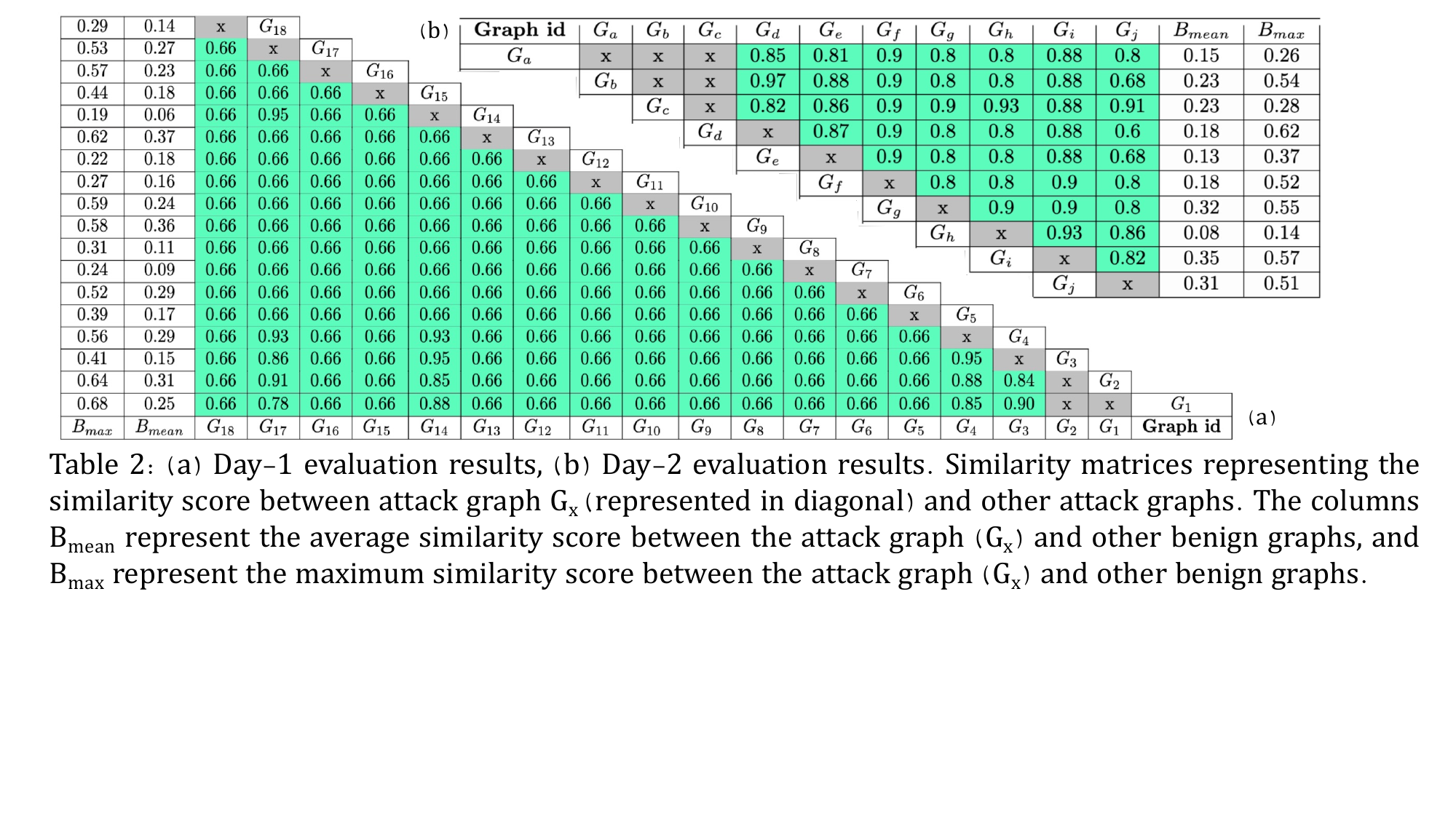}
  \vspace{-22mm}
    \hypertarget{day2}{}
\end{figure*}

Table \hyperlink{day2}{2}(a) represents the results of the first day of activities. During this day, at its "optimal" threshold, \holmes detected only four initial attack graphs from three distinct hosts in its default detection threshold with no false positives. However, the attackers conducted activities in several other hosts. These activities do not cross \holmes' detection threshold. We subsequently reduce \holmes' detection threshold to capture all possible attacker activities. As a result, we generated 424 more graphs from those 500 hosts. \toolname correlated 14 graphs ($G_5$-$G_{18}$) out of those 424 graphs from 14 distinct hosts to be similar to the initial starting points ($G_1$-$G_4$). As can be seen from the table, the average pair-wise similarity values obtained among the attack graphs are significantly higher than those obtained when an attack graph is compared with benign graphs (even with other benign graphs produced from those 14 compromised hosts).

\noindent\textbf{Evaluation on TC dataset.} We evaluated \toolname on one additional dataset generated as part of Engagement-5 organized by DARPA \cite{tc5}. The results are shown in Table \hyperlink{tab:eng5}{3}. The red team used secure ssh sessions to move from one host to another, starting from a pivot host. In each host, the red team performed some suspicious operations (like {\tt nmap, ls, ifconfig}) and exfiltrated a file ({\tt passwd}) back to the pivot machine before moving to a new host. Cadets-1 was a FreeBSD machine in this setup, Theia-1 and Trace-2 were Linux machines, and FiveD-3 was a windows machine. Although the attacks on these different hosts were not identical, due to similar attacker-created processes and invocations across the 4 hosts of different platforms, \toolname was able to detect the similarity across hosts successfully.

\begin{figure}[t]
    \centering
    \includegraphics[width=1\textwidth]{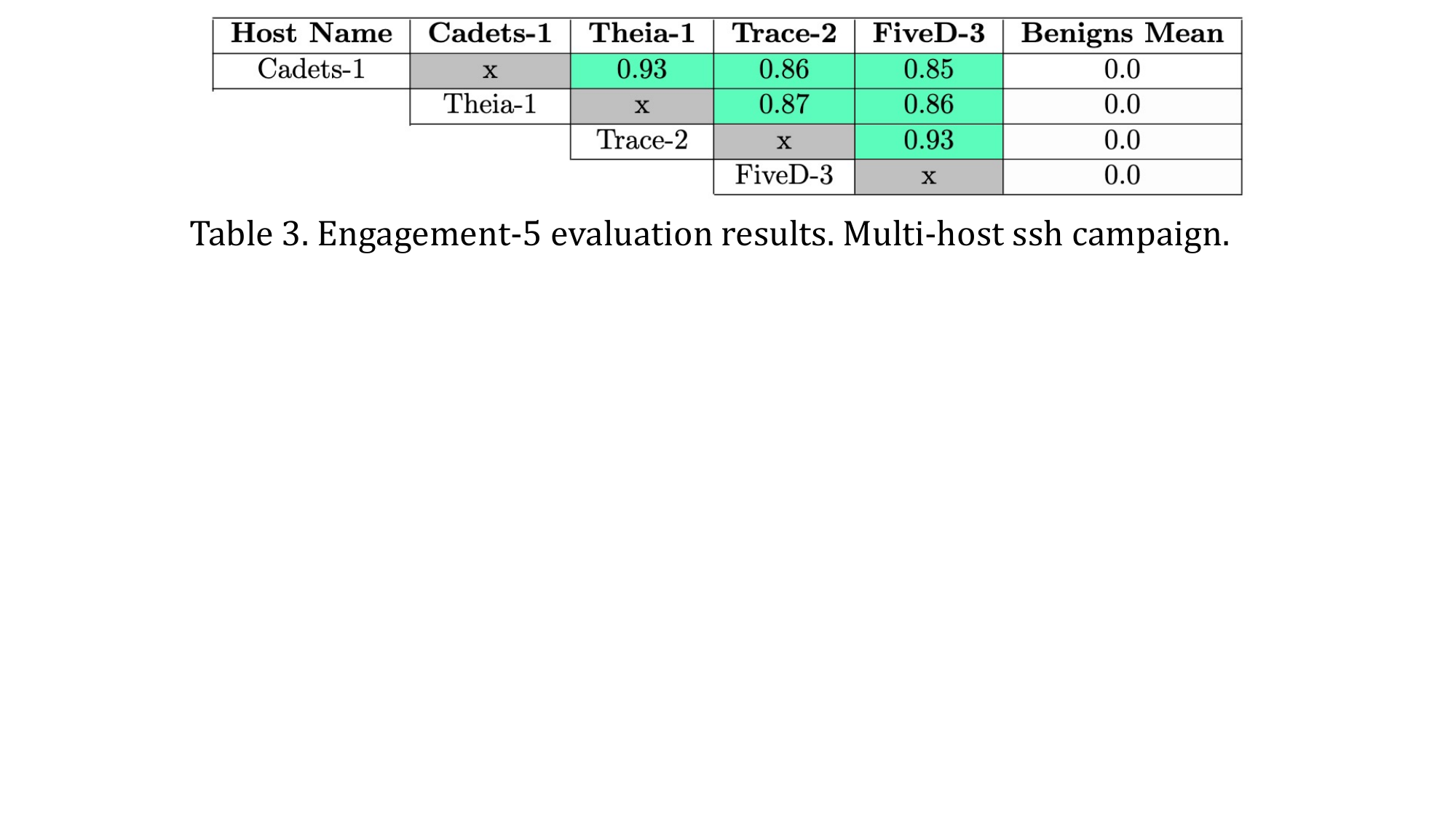}
    \label{tab:eng5}
    \vspace{-49mm}
\end{figure}

\noindent
\textbf{Similarity Threshold, Precision and Recall}.
We define a false positive as a comparison between a benign graph and  an attack graph that results in a value above a specific similarity threshold. In turn, a false negative is a comparison between two attack graphs that results in a value  below the similarity threshold. 
To determine the optimal threshold in our dataset, we varied the threshold over a specific range and collected the false positives, false negatives, true positives, and true negatives using the ground truth. The results of this experiment are shown in Figure \ref{precision_recall_f1}(a), which depicts the values of the precision, recall, F1-score, and accuracy as a function of the similarity threshold. The accuracy metric measures the ratio of correct outcomes over the total number of outcomes ($Accuracy=(TP + TN)/(TP + TN + FP + FN)$). As can be seen in Figure \ref{precision_recall_f1}(a) \toolname achieves a high accuracy ($\sim0.97$),
 Evidently, the optimal value for the F1-score is for values of the similarity threshold around 0.5, which produces a total of 15 false positives over both days.

\subsection{Node Similarity Accuracy}\label{simidet_sec}
In this subsection, we describe an independent evaluation of the approach described in Sections \ref{subsec:tfidf} comparing it with other possible similarity detection methods, in particular, string matching (SM) and \textit{k}-means clustering technique  with a different number of clusters.

\begin{figure*}
    \centering
    \vspace{-2mm}
    \includegraphics[scale=0.38]{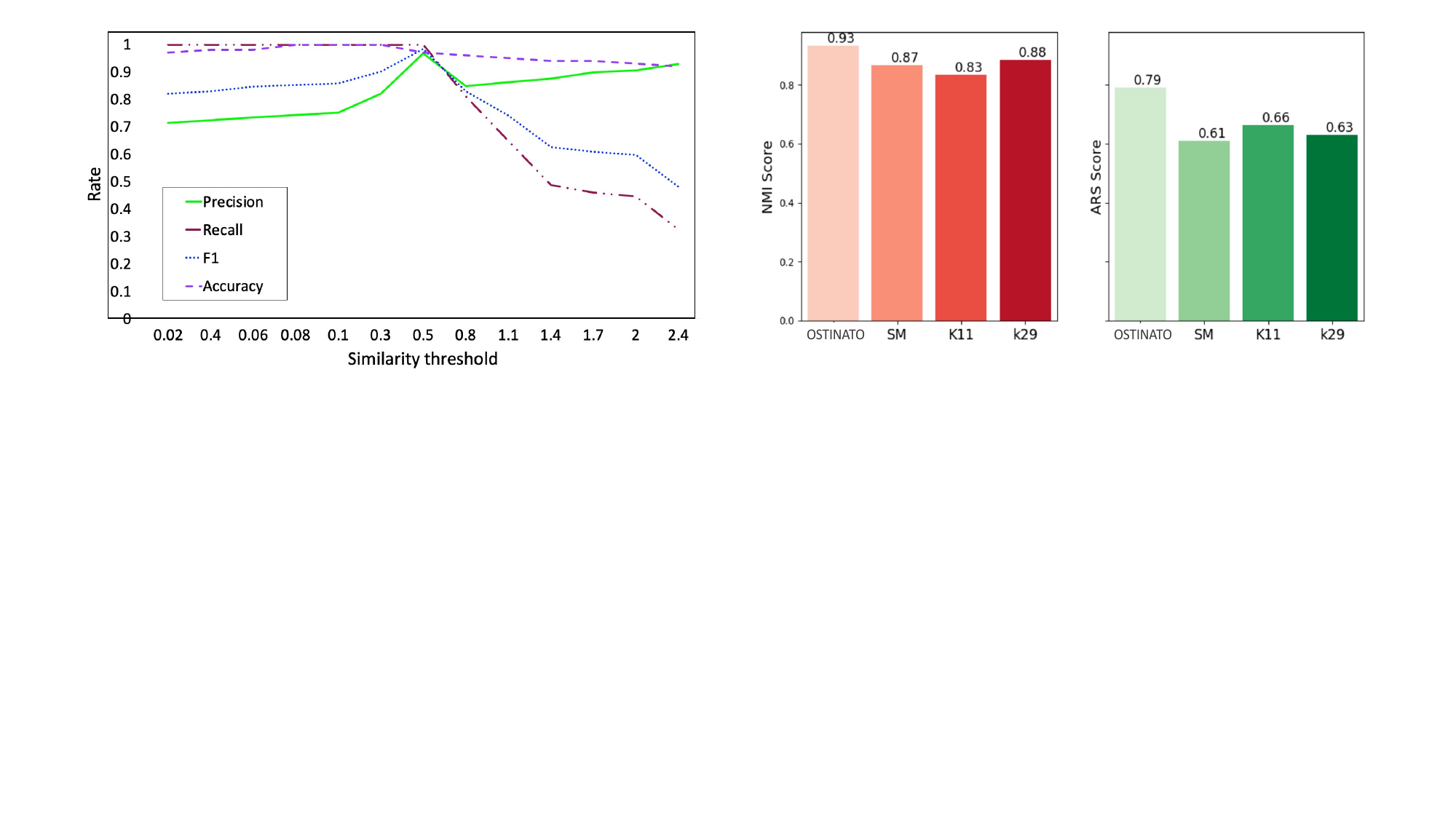}
    \vspace{-43mm}
    \caption{(a) Precision, Recall, F-Score, Accuracy as a function of the similarity detection threshold. (b) Clustering performance comparison between \toolname and other methods using NMI and ARS.}
    \label{precision_recall_f1}
\end{figure*}

\noindent\textbf{Ground-truth Dataset.}
To measure our approach's performance, we created a ground truth dataset of expected buckets for a subset of nodes in the dataset of 3700 subject nodes. We asked multiple security experts to  assign each node to a bucket of similar nodes that represents a specific behavior. After this step, the experts met to discuss their assignments, and if there were disagreements, a new step of assignments were executed. This cycle of assignment-discussions was repeated until consensus was reached.

\noindent\textbf{Comparison Against Other Approaches.} We compare our approach against several common approaches, including SM and the common clustering approach of \textit{k}-means clustering with TF-IDF with different \textit{k} values. To evaluate our approach against the \textit{k}-means clustering, the most fundamental step is to define the optimum number of clusters (i.e., {\emph k}). For this step, we chose two approaches. First, we used the number of clusters based on the number of clusters in the ground truth (i.e., $k = 29$). Second, to choose the number of clusters, we used the elbow method \cite{ketchen1996application}, a common heuristic approach to determine the optimum number of {\emph k} which picks the elbow of the curve as $ k = 11$ as the optimum {\emph k}. Choosing two values for {\emph k} enables us to evaluate our approach against the two probable number of clusters, 1) expected number of clusters based on the ground truth 2) suggested number of (optimum) clusters by elbow method. 

\begin{figure}[b]
    \centering
    \includegraphics[scale=0.38]{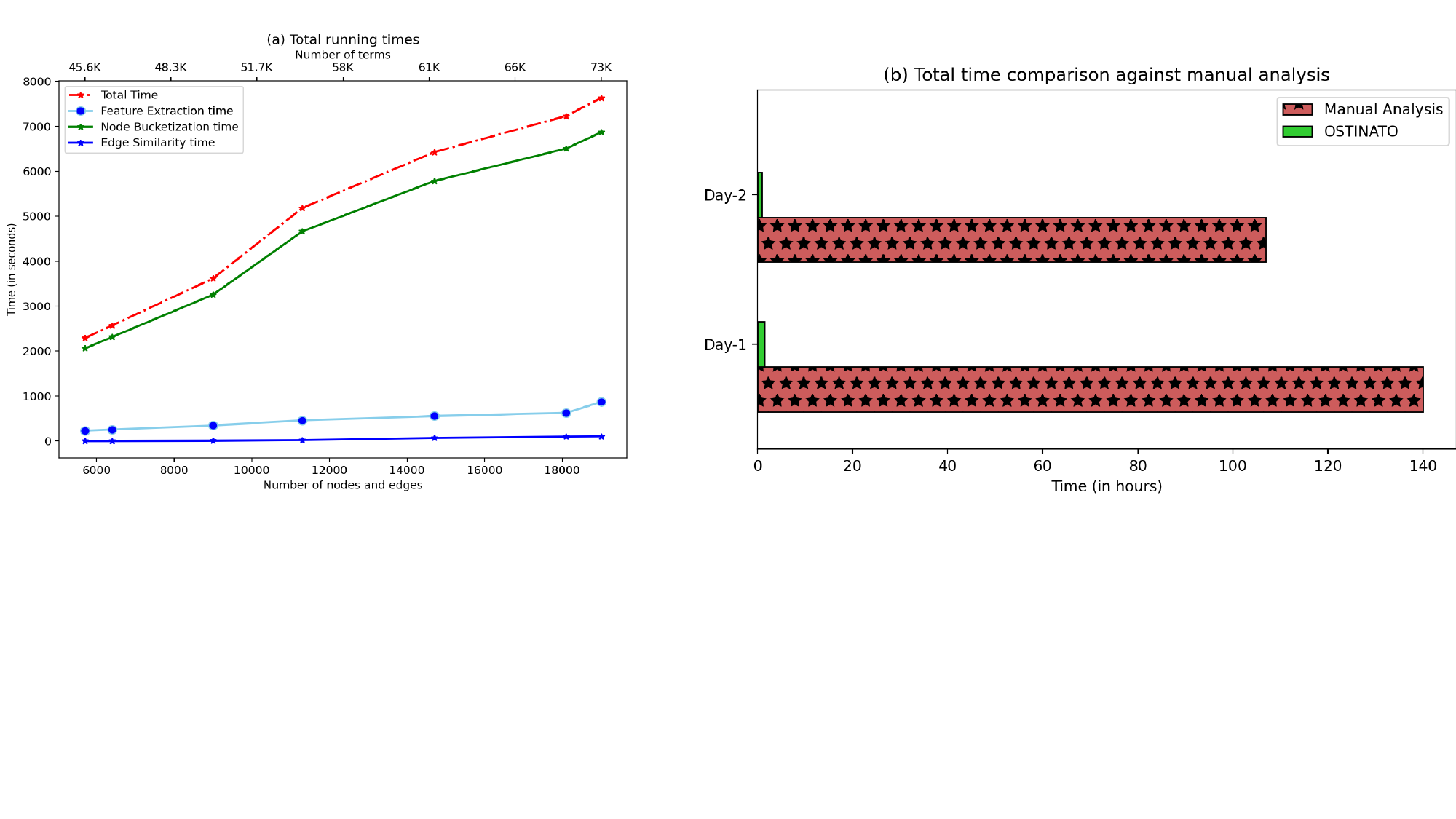}
    \vspace{-35mm}
    \caption{(a) \toolname performance representing the running times of different steps. (b) Total time comparison against manual analysis.}
    \label{fig:time}
\end{figure}

\noindent To measure the performance of our bucketizing approach against other approaches, we use two standard quality metrics for clustering algorithms: the Adjusted Rand Score (ARS) and the Normalized Mutual Information (NMI) metrics \cite{emmons2016analysis} which use different methods to compare the quality of clustering algorithms when the number of clusters in  ground truth clustering and that in the prediction are different. The overall results are shown in Figure \ref{precision_recall_f1}(b). As can be seen from this figure, our approach outperforms both SM and \textit{k}-means for different values of \textit{k}. The main reason for the better performance of our approach is the LSH step, which is able to better capture approximate similarity.

\begin{table}[b]
\renewcommand\thetable{4}
\centering
\begin{adjustbox}{width=0.9\linewidth}
\begin{tabular}{|c|c|c|c|c|} 
\hline
\textbf{Approach} & \textbf{Node label} & \textbf{Node embeddings} & \textbf{Context-relative} & \textbf{Training} \\ 
                  & \textbf{approximation} &                           & \textbf{edge comparison}  & \textbf{required} \\
\hline
 SimGNN \cite{bai2019simgnn}   &         \xmark                    &      \Checkmark                   &  \xmark    &       \Checkmark               \\ 
\hline
 Poirot \cite{milajerdi2019poirot}  &          \xmark                   &        \xmark                 &   \xmark    &     \xmark                 \\ 
\hline
 Deltacon \cite{koutra2013deltacon}  &        \xmark                   &         \Checkmark                &   \xmark    &       \xmark            \\ 
\hline
  \toolname  & \Checkmark                   & \Checkmark                & \Checkmark            &       \xmark  \\
\hline
\end{tabular}
\end{adjustbox}
\caption{\toolname vs. other approaches}
\label{tab:comparison}
\end{table}

\subsection{Run-time performance}
\label{runtime_sec}
We measure the run-time performance of \toolname by creating sets of tagged provenance graphs of different sizes by varying the underlying IDS detection threshold on the {\em Day 1} campaign data. The run time performance of the different steps of \toolname is shown in Figure \ref{fig:time}(a). To obtain different datapoints, we group the graphs into 7 sets of increasing sizes. The number of nodes and edges in each of these sets is shown in the primary x-axis.  The secondary x-axis at the top reflects the total number of words (or terms) present in the nodes of each set.  As can be seen from the figure, the most expensive part of the approach is node bucketization, amounting to approximately 90\% of the total time when comparing  thousands of  nodes. This is mainly due to the large size of the feature matrix. Graph similarity (Algorithm \ref{alg:simi det}), represented by the blue line, is the fastest component, usually taking just a few seconds.

\subsection{Threat Alert Fatigue Mitigation}
 Our evaluation of \toolname for threat alert fatigue mitigation shows promising results. In situations where local host-based detection systems produce a large number of alerts, \toolname can help cyber analysts to pinpoint hosts where similar attacker activities are occurring, filtering out thousands of benign alerts (or false positives from  IDS).  Across the attack campaign \cite{dataset} for two days, the underlying IDS produced more than 1000 alerts, which is really unfeasible for manual analysis. Comparatively, when those graphs are fed into \toolname along with the 7 initial attack graphs, it successfully correlated to 21 alerts from  distinct hosts where it found similar attack behavior. According to several studies \cite{fatigue1,fatigue3,fatigue2},  it usually takes about 10-30 minutes to investigate an alert manually by cyber analysts. Assuming 15 minutes on average for each alarm, a cyber analyst would require 140 hours to investigate the average alarms of each day produced by the IDS in our experiment. Alternatively, as shown in Figure \ref{fig:time}(b), \toolname takes around 167 minutes to complete the analysis of all the alarms generated, reducing  false positives of the underlying IDS by more than 90\%.

\subsection{Comparison with other tools}

\noindent  We compare some of \toolname's aspects with some popular graph matching approaches.  \toolname is much better suited for cross-host attack correlation than compared to other popular graph matching techniques. The features that stand out in comparison with other graph pattern matching approaches is that \toolname can perform accurate node label approximations even when similar nodes exhibit different behaviors, does not require training to implement, and performs context relative edge comparison, which is essential for cross-host attack correlation purposes. We outline the qualitative comparison against the existing tools in Table \ref{tab:comparison}. Since majority of such tools are not open source or easily available, an experimental comparison of \toolname with those approaches is unfeasible. Out of these only SimGNN\cite{bai2019simgnn} is publicly available, however the nodes and edges are much more simpler in it's evaluated datasets and only contains of integers instead of actual names of processes, objects or edges.

%% file: related_works.tex
\section{ Related Work}\label{sec:related}
 
Several approaches have been proposed to deal with cross-host attack detection via cross-host information tracking.
These approaches rely on the presence of information flow data  between entities (e.g., processes) across hosts ~\cite{ji2018enabling,sahabandu2018dift}. This approach, however, requires  fine-grained taint tracking, which relies on system instrumentation it requires some modifications to existing systems.

\noindent\textbf{Log-based Threat Hunting:} A wide variety of systems leverages different types of logs for threat-hunting purposes. Hercule \cite{pei2016hercule} is a log-based detection system modeled on the community discovery problem. It correlates logs from multiple sources and detects attack communities. Oprea et al. \cite{oprea2015detection}, Romero-Gomez et al. \cite{romero2017towards} leveraged DNS, web-proxy logs in order to detect and visualize threats in a network. Bilge et al. \cite{bilge2012disclosure} leveraged NetFlow logs to detect Botnet C\&C servers and distinguish them from the benign traffic. The DNS logs are also leveraged extensively \cite{ antonakakis2012throw} for the detection of malicious domains. Several systems \cite{ liu2018towards,milajerdi2019holmes,hassan2019nodoze} make use of different logs just as \toolname for efficient threat hunting, forensic analysis, or real-time detection of cyberattacks. Most of the mentioned approaches that deal with cross-host activities rely on network logs, however, while \toolname is used over audit host logs.

\noindent \textbf{Provenance Graph Analysis:} BackTracker \cite{king2003backtracking} first introduced the concept of generating a provenance graph from the kernel audit logs. In recent years,  significant progress has been made for log reduction, compression techniques and tracking OS-level dependencies \cite{krishnan2010trail, hossain2020combating, hassan2020tactical} in order to facilitate detection of benign events from the suspicious ones as well as to reduce storage overheads. Moreover, recent studies have used provenance graphs effectively for a wide variety of security problems such as identification of zero-day attack paths in ZePro \cite{sun2018using}, automated provenance triage in NoDoze \cite{hassan2019nodoze}, real-time attack detection and attack scenario reconstruction \cite{milajerdi2019holmes}. While sharing use of provenance graphs, \toolname's approach is different from these works. In fact, \toolname looks for similar subgraphs across multiple provenance graphs as a signal for multi-host attack correlation.

%% file: conclusion.tex
\section{    Conclusion}\label{sec:conclusion}
We present \toolname, which is based on the intuition that attackers have similar goals on multiple hosts during a campaign. \toolname implements an approach for correlating similar attacker activities across different hosts and implements a novel approximate node matching technique. It further uses the attack semantics to detect similarities among tagged provenance graphs. We successfully evaluate \toolname on two datasets created by DARPA red team engagements.